\documentclass{appolb}
\usepackage{epsfig}
\usepackage{amssymb}

\usepackage{prprntno}

\begin{document}
\title{Effects of CP-violating phases in supersymmetry%
\thanks{Presented at the final meeting of the European Network
``Physics at Colliders'', Montpellier, September 26 -- 27, 2004.}%
}
\author{Stefan Hesselbach
\address{Institut f\"ur Theoretische Physik, Universit\"at Wien,
  A-1090 Vienna, Austria}}

\begin{flushright}
\ttfamily
  UWThPh-2004-30 \\
  hep-ph/0410174
\end{flushright}
\vspace{-1.5cm}

\maketitle

\begin{abstract}
Recent studies about the impact of the CP-violating complex parameters
in supersymmetry on the decays of third generation squarks and about
T-odd asymmetries in neutralino and chargino production and decay are
reviewed.
The CP-even branching ratios of the third generation squarks show a
pronounced dependence on the phases of $A_t$, $A_b$, $\mu$ and $M_1$
in a large region of the supersymmetric parameter space.
This could have important implications for stop and sbottom searches
and the MSSM parameter determination in future collider experiments.
We have estimated 
the expected accuracy in the determination of the parameters
by global fits of measured masses, decay branching ratios
and production cross sections. We have found that the parameter
$A_t$ can be determined with an error of 2 -- 3\,\%, whereas the
error on $A_b$ is likely to be of the order of 50 -- 100\,\%.
In addition we have studied
CP-odd observables, like asymmetries based on triple product
correlations, which are necessary to unambiguously establish CP violation.
We have analysed these asymmetries
in neutralino and chargino production with subsequent three-body
decays at the International Linear Collider with longitudinally
polarised beams in the MSSM with complex parameters $M_1$ and $\mu$. 
The asymmetries, which appear already at tree-level because of spin
correlation between production and decay, can be as large as 20\,\%
and will therefore be an important tool for the search for
CP-violating effects in supersymmetry.
\end{abstract}
  
\section{Introduction}

The small amount of CP violation in the Standard Model (SM), which is
caused by the phase in the Cabibbo-Kobayashi-Maskawa matrix, is not
sufficient to explain the baryon-antibaryon asymmetry of the
universe \cite{bau}.
The Lagrangian of the Minimal Supersymmetric Standard Model (MSSM)
contains several complex parameters, which can
give rise to new CP-violating phenomena \cite{mssmcpv}.
In the sfermion sector of the MSSM the trilinear scalar couplings
$A_f$ and the higgsino mass parameter $\mu$ can be complex.
The phase of one of the gaugino mass parameters is
unphysical and can be eliminated.
Conventionally the SU(2) gaugino mass parameter $M_2$ is chosen real, hence
the chargino sector depends only on one complex parameter ($\mu$),
whereas in the neutralino sector also the U(1) gaugino mass parameter
$M_1$ can be complex.

The phases of the complex parameters are constrained or
correlated by the experimental upper limits on the electric
dipole moments of electron, neutron and the atoms
${}^{199}$Hg and ${}^{205}$Tl \cite{edmexp}.
In a constrained MSSM the restrictions on the phases can be rather
severe. However, there
may be cancellations between the contributions of different
complex parameters, which allow larger values for the phases
\cite{edm}.
For example, in mSUGRA-type models and if substantial
cancellations are present, $\varphi_\mu$, the phase of $\mu$, is restricted
to $|\varphi_\mu| \lesssim 0.1\pi$, whereas $\varphi_{M_1}$
and $\varphi_A$, the phases of $M_1$ and the trilinear scalar coupling
parameter, turn out to be essentially unconstrained,
but correlated with $\varphi_\mu$ \cite{Choi:2004rf}.
Moreover, the restrictions are very model dependent.
For example, when also lepton flavour violating terms
are included, then the restriction on $\varphi_\mu$
may disappear \cite{Bartl:2003ju}.

The study of production and decay of 
charginos ($\tilde{\chi}^\pm_i$) and neutralinos ($\tilde{\chi}^0_i$)
and a precise determination of the underlying supersymmetric (SUSY)
parameters $M_1$, $M_2$, $\mu$ and $\tan\beta$
including the phases $\varphi_{M_1}$ and $\varphi_\mu$
will play an important role at the International Linear Collider (ILC)
\cite{LC}.
In \cite{NeuChaParDet} methods to determine these parameters based on
neutralino and chargino mass and cross section measurements have been
presented.
In \cite{Choi:2004rf} the impact of the SUSY phases on chargino,
neutralino and selectron production has been analysed and
significances for the existence of non-vanishing phases have been defined. 
In \cite{Bartl:2004xy} CP-even azimuthal asymmetries in chargino production 
at the ILC with transversely polarised beams have been
analysed. However, CP-odd triple product correlations
involving the transverse beam polarisations vanish, if at least one
subsequent chargino decay is not observed.
Further methods to probe the CP properties of neutralinos 
are described in \cite{Choi:NeutCP}.

In contrast to the parameters of the chargino and neutralino sector
it is more difficult to measure the trilinear couplings $A_f$ in the
sfermion sector.
In the real MSSM studies to determine $A_f$ have been performed in
\cite{realAf}.
In the complex MSSM the polarisation of final top quarks and tau
leptons from the
decays of third generation sfermions can be a sensitive probe of the
CP-violating phases \cite{Gajdosik:2004ed}.
In \cite{staupapers} the effects of the CP phases of $A_\tau$, $\mu$
and $M_1$ on production and decay of tau sleptons ($\tilde{\tau}_{1,2}$) and
tau sneutrinos ($\tilde{\nu}_\tau$) have been studied.
The branching ratios of $\tilde{\tau}_{1,2}$ and $\tilde{\nu}_\tau$
can show a strong phase dependence. The expected accuracy in the
determination of $A_\tau$ has been estimated to be 10\,\% by a global
fit of measured masses, branching ratios and production cross sections.
The impact of the SUSY phases on the decays of the
third generation squarks \cite{squarkpapers, Bartl:2003pd} will be
discussed in Sec.~\ref{secSquarkdecay}.

However, in order to unambiguously establish CP violation in supersymmetry,
including the signs of the phases, the use of CP-odd
observables is inevitable.
In supersymmetry T-odd triple product correlations between momenta and
spins of the
involved particles allow the definition of CP-odd asymmetries already
at tree level \cite{Choi:1999cc, tripleproducts}.
Such asymmetries based on triple products in decays of scalar fermions
have been discussed in \cite{ATsfermion}.
T-odd asymmetries in neutralino and chargino production with
subsequent two-body decays have been analysed in \cite{AT2body}.
For leptonic two-body decays asymmetries up to 30\,\% can occur.
CP-odd observables involving
the polarisation of final $\tau$ leptons from two-body decays of neutralinos 
have been studied in \cite{Ataupol}.
A Monte Carlo study of T-odd asymmetries in selectron and neutralino
production and decay including initial state radiation,
beamstrahlung, SM backgrounds and detector effects
has been given in \cite{MonteCarlo}.
It has been found that asymmetries $\mathcal{O}(10\,\%)$ are detectable
after few years of running of the ILC.
T-odd asymmetries in neutralino and chargino production with
subsequent three-body decays \cite{Bartl:2004jj, charginopaper}
will be discussed in Sec.~\ref{SecAT}.

\section{\label{secSquarkdecay} Decays of third generation squarks}

In \cite{squarkpapers, Bartl:2003pd} we have studied the effects of
the phases of
the parameters $A_t$, $A_b$, $\mu$ and $M_1$ on the phenomenology of the
third generation squarks, the stops $\tilde{t}_{1,2}$ and the sbottoms
$\tilde{b}_{1,2}$ in the complex MSSM.
We have focused especially on the effects of $\varphi_{A_t}$ and
$\varphi_{A_b}$ in order to find possibilities to determine these
parameters.
The third generation squark sector is particularly interesting because
of the effects of the large Yukawa couplings.
The phases of $A_f$ and $\mu$ enter directly the squark mass
matrices and the squark-Higgs couplings, which can cause a strong
phase dependence of observables.
The off-diagonal mass matrix element $M^2_{\tilde{q}RL}$, which
describes the mixing between the left and right squark states, is
given by
\begin{equation}
 M^2_{\tilde{t}RL} = m_t \left( |A_t| e^{i\varphi_{A_t}} - 
   \frac{|\mu| e^{-i\varphi_{\mu}}}{\tan\beta} \right),
\end{equation}
\begin{equation}
 M^2_{\tilde{b}RL} = m_b \left( |A_b| e^{i\varphi_{A_b}} - 
   |\mu| e^{-i\varphi_{\mu}} \tan\beta \right)
\end{equation}
for the stops and sbottoms, respectively.
In the case of stops the $\mu$ term is suppressed by $1/\tan\beta$,
hence the phase $\varphi_{\tilde{t}}$ of $M^2_{\tilde{t}RL}$ is
dominated by $\varphi_{A_t}$, 
in a large part of the SUSY parameter space with $|A_t| \gg |\mu|/\tan\beta$.
$\varphi_{\tilde{t}} \approx \varphi_{A_t}$ enters the stop mixing matrix
and in the following all couplings because of the strong mixing in the
stop sector. This can lead to a strong phase dependence of many
partial decay widths and branching ratios.

In the case of sbottoms the mixing is smaller because of the
smaller bottom mass. It is mainly important for large
$\tan\beta$, when the $\mu$ term is dominant in
$M^2_{\tilde{b}RL}$. Hence the phase of $A_b$ has only minor impact on
the sbottom mixing in a large part of the SUSY parameter space.
However, in the squark-Higgs couplings, for example in the 
$H^\pm \tilde{t}_L \tilde{b}_R$ coupling
\begin{equation} \label{eq:ctHb}
C(\tilde{t}_L^\dagger H^+ \tilde{b}_R) \sim m_b
  \left( |A_b| e^{-i\varphi_{A_b}} \tan\beta + |\mu|
    e^{i\varphi_{\mu}} \right),
\end{equation}
the phase $\varphi_{A_b}$ appears independent of the sbottom mixing.
This can lead to a strong $\varphi_{A_b}$ dependence of sbottom
\emph{and} stop partial decay widths into Higgs bosons.

\subsection{Partial decay widths and branching ratios of stops and sbottoms}

In this subsection we discuss the $\varphi_{A_t}$ and $\varphi_{A_b}$
dependence of stop and sbottom partial decay widths and branching
ratios. We have analysed fermionic decays 
$\tilde{q}_i \to \tilde{\chi}^\pm_j q'$,
  $\tilde{q}_i \to \tilde{\chi}^0_j q$
and bosonic decays
$\tilde{q}_i \to \tilde{q}_j' H^\pm$,
  $\tilde{q}_i \to \tilde{q}_j' W^\pm$,
  $\tilde{q}_2 \to \tilde{q}_1 H_i$,
  $\tilde{q}_2 \to \tilde{q}_1 Z$
of $\tilde{t}_{1,2}$ and $\tilde{b}_{1,2}$.
In the complex MSSM the CP-even and CP-odd neutral Higgs bosons mix
and form three mass eigenstates $H_{1,2,3}$.
Their masses and mixing matrices have been
calculated with the program FeynHiggs2.0.2 \cite{FeynHiggs}.

\begin{figure}[t]
\centerline{\epsfig{file=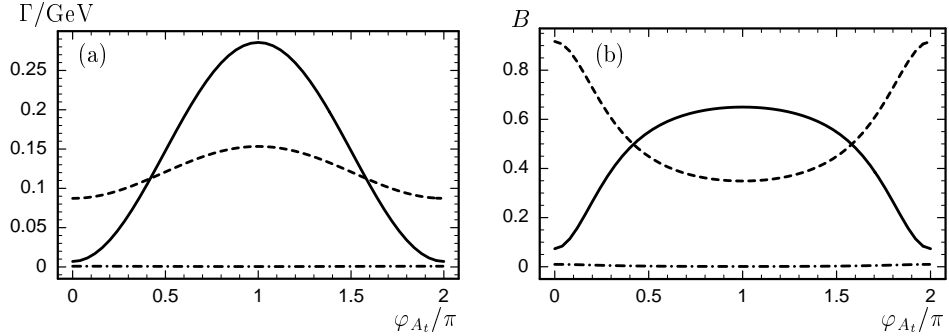,scale=0.78}}
\caption{\label{fig:stop1decays}
(a) Partial decay widths $\Gamma$
and (b) branching ratios $B$ of the decays
$\tilde{t}_1 \to \tilde{\chi}^+_1 b$ (solid),
$\tilde{t}_1 \to \tilde{\chi}^0_1 t$ (dashed) and
$\tilde{t}_1 \to W^+ \tilde{b}_1$ (dashdotted)
for $\tan\beta = 6$, $M_2=300$~GeV,
$|M_1|/M_2 = 5/3 \, \tan^2\theta_W$, $|\mu|=350$~GeV,
$|A_b|=|A_t|=800$~GeV,
$\varphi_\mu=\pi$, $\varphi_{M_1}=\varphi_{A_b}=0$,
$m_{\tilde{t}_1}=350$~GeV, $m_{\tilde{t}_2}=700$~GeV,
$m_{\tilde{b}_1}=170$~GeV, $M_{\tilde{Q}}>M_{\tilde{U}}$
and $m_{H^\pm}=900$~GeV.
From \protect\cite{Bartl:2003pd}.}
\end{figure}

\begin{figure}[t]
\centerline{\epsfig{file=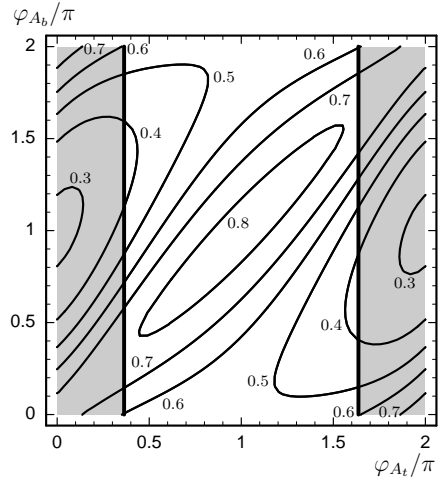,scale=.75}}
\caption{\label{fig:cpcharg}Contours of
$B(\tilde{t}_1 \to \tilde{\chi}^+_1 b)$
for $\tan\beta = 30$, $M_2=300$~GeV, $|\mu|=300$~GeV,
$|A_b|=|A_t|=600$~GeV, $\varphi_\mu=\pi$, $\varphi_{M_1}=0$,
$m_{\tilde{t}_1}=350$~GeV, $m_{\tilde{t}_2}=700$~GeV,
$m_{\tilde{b}_1}=170$~GeV,
$M_{\tilde{Q}}>M_{\tilde{U}}$
and $m_{H^\pm}=160$~GeV.
The shaded areas are excluded by the experimental limit
$B(b\to s \gamma) > 2.0 \times 10^{-4}$.
From \cite{Bartl:2003pd}.}
\end{figure}

In the scenario of Fig.~\ref{fig:stop1decays}, where
we show the partial decay widths
$\Gamma$ and branching ratios $B$ of the $\tilde{t}_1$ decays,
especially $\Gamma(\tilde{t}_1 \to \tilde{\chi}^+_1 b)$ 
has a very pronounced $\varphi_{A_t}$ dependence, which leads to a
strong $\varphi_{A_t}$ dependence of the branching ratios.
This pronounced $\varphi_{A_t}$ dependence of 
$\Gamma(\tilde{t}_1 \to \tilde{\chi}^+_1 b)$ is caused by the phase 
$\varphi_{\tilde{t}} \approx \varphi_{A_t}$ of the stop mixing matrix
which enters the respective couplings.
In the scenario of Fig.~\ref{fig:cpcharg} with large $\tan\beta$ the decay channel
$\tilde{t}_1 \to H^+ \tilde{b}_1$ is open.
$\varphi_{A_b}$ in the $H^\pm \tilde{t}_L \tilde{b}_R$ coupling,
Eq.~(\ref{eq:ctHb}), causes a strong $\varphi_{A_b}$ dependence of
$\Gamma(\tilde{t}_1 \to H^+ \tilde{b}_1)$, which influences all
branching ratios. An interplay between the $\varphi_{A_t}$ dependence
of the stop mixing matrix and the $\varphi_{A_b}$ dependence of the 
$H^\pm \tilde{t}_L \tilde{b}_R$ coupling leads to the
remarkable correlation between $\varphi_{A_b}$ and $\varphi_{A_t}$.
In the case of the heavy $\tilde{t}_2$ many decay channels
can be open and can show a strong $\varphi_{A_t}$ dependence.

\begin{figure}[t]
\centerline{\epsfig{file=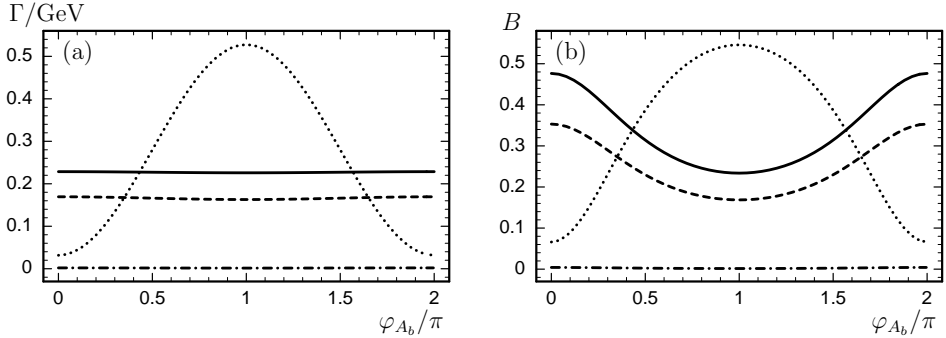,scale=0.78}}
\caption{\label{fig:sbottom1decays}
(a) Partial decay widths $\Gamma$ and (b) branching
ratios $B$ of the decays
$\tilde{b}_1 \to \tilde{\chi}^0_1 b$ (solid),
$\tilde{b}_1 \to \tilde{\chi}^0_2 b$ (dashed),
$\tilde{b}_1 \to H^- \tilde{t}_1$ (dotted) and
$\tilde{b}_1 \to W^- \tilde{t}_1$ (dashdotted)
for $\tan\beta = 30$, $M_2=200$~GeV, $|\mu| = 300$~GeV, $|A_b|=|A_t|=600$~GeV,
$\varphi_\mu=\pi$, $\varphi_{A_t}=\varphi_{M_1}=0$,
$m_{\tilde{b}_1}=350$~GeV, $m_{\tilde{b}_2}=700$~GeV,
$m_{\tilde{t}_1}=170$~GeV,
$M_{\tilde{Q}}>M_{\tilde{D}}$
and $m_{H^\pm}=150$~GeV.
From \cite{Bartl:2003pd}.}
\end{figure}

In Fig.~\ref{fig:sbottom1decays},
where $\tilde{b}_1$ decays are discussed, only
$\Gamma(\tilde{b}_1 \to H^- \tilde{t}_1)$ shows a pronounced $\varphi_{A_b}$
dependence, which leads to a strong $\varphi_{A_b}$ dependence of the
branching ratios. This is again caused by $\varphi_{A_b}$ entering the 
$H^\pm \tilde{t}_L \tilde{b}_R$ coupling Eq.~(\ref{eq:ctHb}).
The other partial decay widths depend only very weakly on
$\varphi_{A_b}$.
This is typical for the $\varphi_{A_b}$ dependence of the $\tilde{b}_{1,2}$
decays.
Only the partial decay widths into Higgs bosons 
($\Gamma(\tilde{b}_1 \to H^- \tilde{t}_1)$ for $\tilde{b}_1$ and
$\Gamma(\tilde{b}_2 \to H^- \tilde{t}_{1,2})$, 
$\Gamma(\tilde{b}_2 \to H_{1,2,3} \tilde{b}_1)$ for $\tilde{b}_2$)
can show a strong phase dependence for large $\tan\beta$.

\subsection{Parameter determination via global fit}

In order to estimate the precision, which can be expected in the
determination of the underlying SUSY parameters, we have made a global
fit of many observables in \cite{Bartl:2003pd}.
In order to achieve this the following assumptions have been made:
(i) At the ILC the masses of the charginos, neutralinos and the
lightest Higgs boson can be measured with high precision.
If the masses of the squarks and heavier Higgs bosons are below
500~GeV, they can be measured with an error of $1\,\%$ and 1.5~GeV,
respectively.
(ii) The masses of the squarks and heavier Higgs bosons, which are
heavier than 500~GeV, can be measured at a 2~TeV $e^+ e^-$ collider
like CLIC with an error of $3\,\%$ and $1\,\%$, respectively.
(iii) The gluino mass can be measured at the LHC with an error of
$3\,\%$.
(iv) For the production cross sections 
$\sigma(e^+ e^- \to \tilde{t}_i \bar{\tilde{t}}_j)$ and
$\sigma(e^+ e^- \to \tilde{b}_i \bar{\tilde{b}}_j)$ and
the branching ratios of the $\tilde{t}_i$ and $\tilde{b}_i$ decays
we have taken the statistical errors, which we have doubled to be on
the conservative side.
We have analysed two scenarios, one with small $\tan\beta = 6$ and one with
large $\tan\beta = 30$.
In both scenarios we have found that $\mathrm{Re}(A_t)$ and
$|\mathrm{Im}(A_t)|$ can be determined with relative errors of 2 -- $3\,\%$.
For $A_b$ the situation is considerably worse because of the weaker
dependence of the observables on this parameter. Here the
corresponding errors are of the order of 50 -- 100\,\%.
For the squark mass parameters
$M_{\tilde{Q}}$, $M_{\tilde{U}}$, $M_{\tilde{D}}$
the relative errors are of order of 1\,\%, for $\tan\beta$ of order of
3\,\% and for the other fundamental SUSY parameters of order of 1 -- 2\,\%.

\section{\label{SecAT}T-odd asymmetries in neutralino and chargino
  production and decay}

We have studied T-odd asymmetries
in neutralino \cite{Bartl:2004jj} and chargino \cite{charginopaper}
production with subsequent three-body decays
\begin{equation} \label{ncprocess}
e^+ e^- \to \tilde{\chi}_i + \tilde{\chi}_j \to
\tilde{\chi}_i + \tilde{\chi}^0_1 f \bar{f}^{(')},
\end{equation}
where full
spin correlations between production and decay have to be included
\cite{spincorr}.
Then in the amplitude squared $|T|^2$ of the combined process products like
$i\epsilon_{\mu\nu\rho\sigma}p^\mu_ip^\nu_jp^\rho_kp^\sigma_l$,
where the $p^\mu_i$ denote the momenta of the involved particles,
appear in the terms, which depend on the spin of the decaying neutralino
or chargino. Together with the complex couplings these terms can give
real contributions to suitable observables at tree-level.
Triple products
$\mathcal{T}_1 = \vec{p}_{e^-}\cdot(\vec{p}_{f}\times\vec{p}_{\bar{f}^{(')}})$
of the initial electron momentum $\vec{p}_{e^-}$ and
the two final fermion momenta $\vec{p}_{f}$ and $\vec{p}_{\bar{f}^{(')}}$
or
$\mathcal{T}_2 = \vec{p}_{e^-}\cdot(\vec{p}_{\tilde{\chi}_j}\times\vec{p}_{f})$
of the initial electron momentum $\vec{p}_{e^-}$, the momentum of the
decaying neutralino or chargino $\vec{p}_{\tilde{\chi}_j}$ and one
final fermion momentum $\vec{p}_{f}$
allow the definition of T-odd asymmetries
\begin{equation}
A_T = \frac{\sigma({\cal T}_i>0) - \sigma({\cal T}_i<0)}%
 {\sigma({\cal T}_i>0) + \sigma({\cal T}_i<0)}
 =
 \frac{\int {\rm sign}({\cal T}_i) |T|^2 d{\rm Lips}}%
 {{\int}|T|^2 d{\rm Lips}},
\end{equation}
where ${\int}|T|^2 d{\rm Lips}$ is
proportional to the cross section $\sigma$ of the process (\ref{ncprocess}).
$A_T$ is odd under naive time-reversal operation and hence CP-odd, if
higher order
final-state interactions and finite-widths effects can be neglected.

\subsection{T-odd asymmetry in neutralino production and decay}

\begin{figure}[t]
\centerline{\epsfig{file=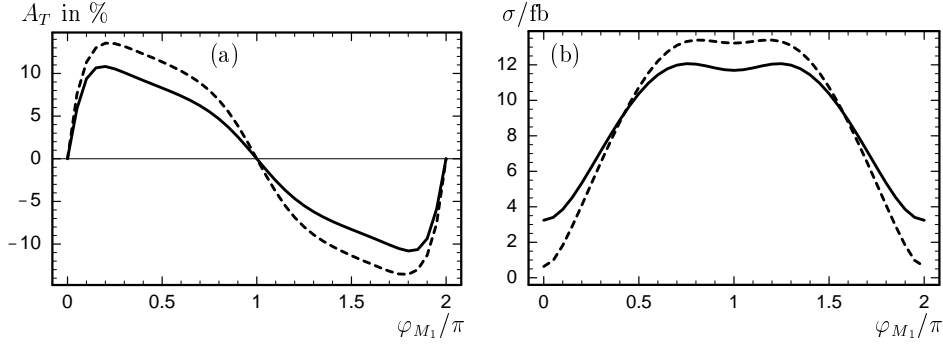,scale=0.78}}
\caption{\label{fig:At12}
(a) T-odd asymmetry $A_T$ and (b) cross section
$\sigma(e^+e^- \to \tilde{\chi}^0_1\tilde{\chi}^0_2 \to$
$\tilde{\chi}^0_1\tilde{\chi}^0_1 \ell^+ \ell^-)$,
summed over $\ell=e,\mu\,$,
for $|M_1|=150$~GeV, $M_2=300$~GeV, $|\mu|=200$~GeV, $\tan\beta=10$,
$m_{\tilde{\ell}_L}=267.6$~GeV, $m_{\tilde{\ell}_R}=224.4$~GeV
and $\varphi_{\mu}=0$ at the ILC with beam polarisations
$P_{e^-}=-0.8$, $P_{e^+}=+0.6$ and
$\sqrt{s}=500$~GeV (solid), $\sqrt{s}=350$~GeV (dashed).
From \cite{Bartl:2004jj}.}
\end{figure}

\begin{figure}[t]
\centerline{\epsfig{file=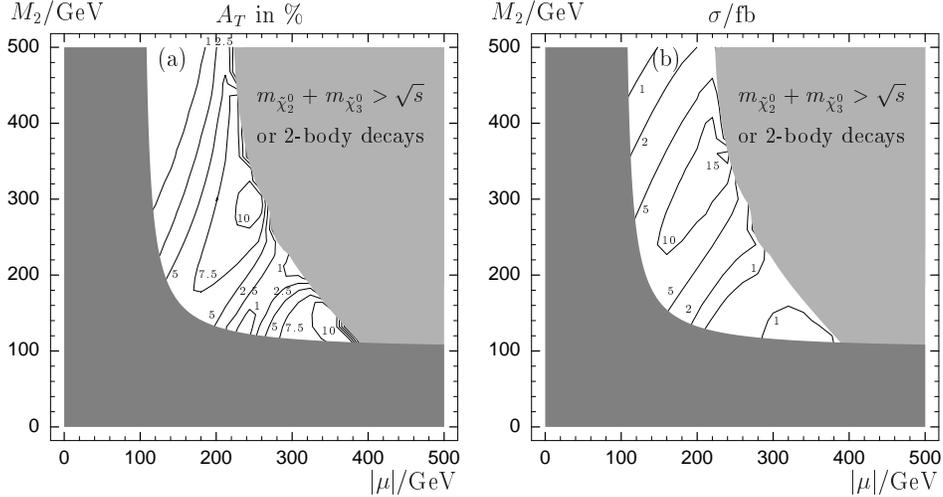,scale=0.78}}
\caption{\label{fig:At32}
Contours (a) of the T-odd asymmetry $A_T$ in \%
and (b) of the cross section
$\sigma(e^+e^- \to \tilde{\chi}^0_3\tilde{\chi}^0_2 \to$
$\tilde{\chi}^0_3\tilde{\chi}^0_1 \ell^+ \ell^-)$,
summed over $\ell=e,\mu\,$,
in fb, respectively,
for $\tan\beta = 10$, $m_{\tilde{\ell}_L} = 267.6$~GeV,
$m_{\tilde{\ell}_R} = 224.4$~GeV, $|M_1|/M_2 = 5/3 \tan^2\theta_W$,
$\varphi_{M_1}=0.5\pi$ and $\varphi_{\mu}=0$
with $\sqrt{s}=500$~GeV and $P_{e^-}=-0.8$, $P_{e^+}=+0.6$.
The dark shaded area marks the parameter space with
$m_{\tilde{\chi}^\pm_1} < 103.5$~GeV excluded by LEP.
The light shaded area is kinematically not accessible or
in this area the analysed three-body decay is strongly
suppressed because $m_{\tilde{\chi}^0_2} > m_Z + m_{\tilde{\chi}^0_1}$
or $m_{\tilde{\chi}^0_2} > m_{\tilde{\ell}_R}$, respectively.
From \cite{Bartl:2004jj}.}
\end{figure}

In neutralino production and subsequent leptonic three-body decay 
$e^+ e^-$ $\to \tilde{\chi}^0_i + \tilde{\chi}^0_2 \to
\tilde{\chi}^0_i + \tilde{\chi}^0_1 \ell^+ \ell^-$
the triple product 
$\mathcal{T}_1 = \vec{p}_{e^-}\cdot(\vec{p}_{\ell^+}\times\vec{p}_{\ell^-})$
can be used to define $A_T$. Then $A_T$ can be directly measured without
reconstruction of the momentum of the decaying neutralino or further
final-state analyses.
As can be seen in Fig.~\ref{fig:At12},
asymmetries $A_T = \mathcal{O}(10\,\%)$ are possible with corresponding
cross sections $\sigma = \mathcal{O}(10~\mathrm{fb})$
for the associated production and decay of $\tilde{\chi}^0_1$ and
$\tilde{\chi}^0_2$,
$e^+ e^- \to \tilde{\chi}^0_1 + \tilde{\chi}^0_2 \to
\tilde{\chi}^0_1 + \tilde{\chi}^0_1 \ell^+ \ell^-$.
For a centre of mass energy $\sqrt{s} = 350$~GeV, i.e.\ closer to
threshold of production, $A_T$ is larger, which is typical for
effects caused by spin correlations between production and decay.
Also for the associated production and decay of $\tilde{\chi}^0_2$ and
$\tilde{\chi}^0_3$,
$e^+ e^- \to \tilde{\chi}^0_3 + \tilde{\chi}^0_2 \to
\tilde{\chi}^0_3 + \tilde{\chi}^0_1 \ell^+ \ell^-$,
the asymmetry $A_T$ has values $\mathcal{O}(10\,\%)$ in large
parameter regions, where the corresponding cross sections 
$\sigma = \mathcal{O}(10~\mathrm{fb})$ (see Fig.~\ref{fig:At32}).
However, for
$e^+ e^- \to \tilde{\chi}^0_2 + \tilde{\chi}^0_2 \to
\tilde{\chi}^0_2 + \tilde{\chi}^0_1 \ell^+ \ell^-$
all couplings in the $\tilde{\chi}^0_2$ pair
production process are real, which leads to small 
asymmetries $A_T = \mathcal{O}(1\,\%)$,
whereas for
$e^+ e^- \to \tilde{\chi}^0_4 + \tilde{\chi}^0_2 \to
\tilde{\chi}^0_4 + \tilde{\chi}^0_1 \ell^+ \ell^-$
with larger asymmetries $A_T \approx 6\,\%$
the cross section $\sigma \lesssim 1~\mathrm{fb}$ is rather small.

\subsection{T-odd asymmetry in chargino production and decay}

\begin{figure}[t]
\centerline{\epsfig{file=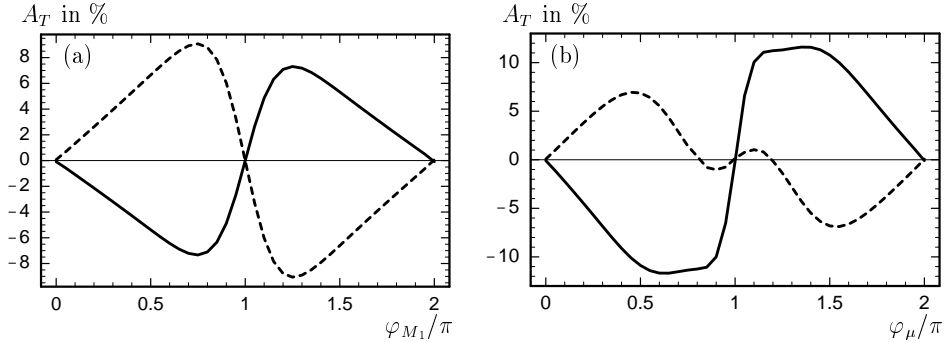,scale=0.78}}
\caption{\label{fig:Atchar12}
T-odd asymmetry $A_T$
for $e^+e^- \to \tilde{\chi}^-_2 + \tilde{\chi}^+_1
 \to \tilde{\chi}^-_2 + \tilde{\chi}^0_1 \bar{s} c$
in the scenario
$M_2=150$~GeV, $|M_1|/M_2= 5/3 \tan^2\theta_W$, $|\mu|=320$~GeV,
$\tan\beta=5$, $m_{\tilde{\nu}}=250$~GeV and $m_{\tilde{u}_L}=500$~GeV
with $\sqrt{s} = 500$~GeV
(a) for $\varphi_{\mu}=0$ and (b) for $\varphi_{M_1}=0$ and
beam polarisations
$P_{e^-}=-0.8$, $P_{e^+}=+0.6$ (solid),
$P_{e^-}=+0.8$, $P_{e^+}=-0.6$ (dashed).}
\end{figure}

In chargino production and subsequent hadronic three-body decay 
$e^+ e^-$ $\to \tilde{\chi}^-_i + \tilde{\chi}^+_1 \to
\tilde{\chi}^-_i + \tilde{\chi}^0_1 \bar{s} c$
the triple product 
$\mathcal{T}_1 = \vec{p}_{e^-}\cdot(\vec{p}_{\bar{s}}\times\vec{p}_{c})$
can be used to define $A_T$.
In this case it is important to tag the $c$ jet to discriminate
between the two jets and to measure the sign of $\mathcal{T}_1$.
For the associated production and decay of $\tilde{\chi}^+_1$ and
$\tilde{\chi}^-_2$,
$e^+ e^-$ $\to \tilde{\chi}^-_2 + \tilde{\chi}^+_1 \to
\tilde{\chi}^-_2 + \tilde{\chi}^0_1 \bar{s} c$,
asymmetries $A_T = \mathcal{O}(10\,\%)$ are possible
(Fig.~\ref{fig:Atchar12}).
In the scenario of Fig.~\ref{fig:Atchar12} the corresponding cross
sections are in the range of 1 -- 5~fb.
In Fig.~\ref{fig:Atchar12} (b) it is remarkable that large asymmetries
$A_T \approx 10\,\%$ are reached for small complex $\varphi_\mu$
around $\varphi_\mu = \pi$.
In the chargino sector even for the pair production and decay process
$e^+ e^-$ $\to \tilde{\chi}^-_1 + \tilde{\chi}^+_1 \to
\tilde{\chi}^-_1 + \tilde{\chi}^0_1 \bar{s} c$
asymmetries $A_T \approx 5\,\%$ can appear, which can only originate
from the decay process. This means that the contributions from the
decay to $A_T$ play an important role in chargino production with
subsequent hadronic decays, which can can also be seen in 
Fig.~\ref{fig:Atchar12} (a) with large $A_T$ for 
$\varphi_\mu = 0$ and $\varphi_{M_1} \neq 0$.
It is furthermore remarkable that 
$\sigma(e^+ e^-$ $\to \tilde{\chi}^-_1 + \tilde{\chi}^+_1 \to
\tilde{\chi}^-_1 + \tilde{\chi}^0_1 \bar{s} c)$
can be rather large, for example 117~fb in the scenario $M_2 = 350$~GeV, 
$|\mu| = 260$~GeV and the other parameters as in 
Fig.~\ref{fig:Atchar12} (a), where $A_T \approx 4\,\%$.

\begin{figure}[t]
\centerline{\epsfig{file=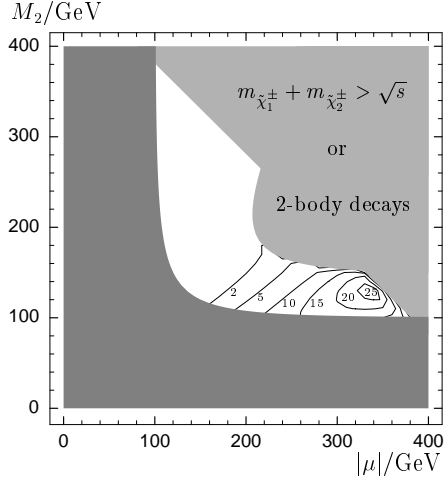,scale=0.75}}
\caption{\label{fig:Atchar12L}
Contours of the T-odd asymmetry $A_T$ in \% for
$e^+e^- \to \tilde{\chi}^-_2 + \tilde{\chi}^+_1
 \to \tilde{\chi}^-_2 + \tilde{\chi}^0_1 \ell^+ \nu$
for
$\varphi_{\mu}=0.9\pi$, $\varphi_{M_1}=1.5\pi$,
$|M_1|/M_2= 5/3 \tan^2\theta_W$, 
$\tan\beta=5$ and $m_{\tilde{\nu}}=250$~GeV
with $\sqrt{s} = 500$~GeV and $P_{e^-}=+0.8$, $P_{e^+}=-0.6$.
The dark shaded area marks the parameter space with
$m_{\tilde{\chi}^\pm_1} < 103.5$~GeV excluded by LEP.
The light shaded area is kinematically not accessible or
in this area the analysed three-body decay is strongly
suppressed because $m_{\tilde{\chi}^\pm_1} > m_W + m_{\tilde{\chi}^0_1}$,
respectively.
}
\end{figure}

If the momentum of the decaying chargino $\tilde{\chi}^+_1$ can be
reconstructed, for example with help of information from the decay of the
$\tilde{\chi}^-_i$, 
the process
$e^+e^- \to \tilde{\chi}^-_i + \tilde{\chi}^+_1
 \to \tilde{\chi}^-_i + \tilde{\chi}^0_1 \ell^+ \nu$
can be analysed, where the chargino decays leptonically.
Then the triple product 
$\mathcal{T}_2 = \vec{p}_{e^-}\cdot(\vec{p}_{\tilde{\chi}^+_1}\times\vec{p}_{\ell^+})$
can be used to define $A_T$.
For the associated production and decay of $\tilde{\chi}^+_1$ and
$\tilde{\chi}^-_2$,
$e^+ e^-$ $\to \tilde{\chi}^-_2 + \tilde{\chi}^+_1 \to
\tilde{\chi}^-_2 + \tilde{\chi}^0_1 \ell^+ \nu$,
asymmetries $A_T \gtrsim 20\,\%$ can occur (Fig.~\ref{fig:Atchar12L}).
But in the region with largest asymmetries around
$|\mu| = 320$~GeV and $M_2 = 120$~GeV the cross section is very
small ($\sigma = \mathcal{O}(0.1~\mathrm{fb})$).
However, for decreasing $|\mu|$ the cross section increases and
reaches $\sigma = 2$~fb for $|\mu| = 220$~GeV and $M_2 = 120$~GeV.
For pair production of $\tilde{\chi}^\pm_1$ and leptonic decays,
$e^+ e^-$ $\to \tilde{\chi}^-_1 + \tilde{\chi}^+_1 \to
\tilde{\chi}^-_1 + \tilde{\chi}^0_1 \ell^+ \nu$,
the couplings in the production process are real which leads to small
$A_T = \mathcal{O}(1\,\%)$, because the contributions from the decay
are not large enough in this case.

\section{Conclusions}

We have studied the impact of the complex parameters $A_t$, $A_b$,
$\mu$ and $M_1$ on the decays of stops and sbottoms in the
CP-violating MSSM.
In the case of stop decays all partial decay widths and branching
ratios can have a strong $\varphi_{A_t}$ dependence because of the large
mixing in the stop sector.
If $\tan\beta$ is large and decay channels into Higgs bosons are open,
stop \emph{and} sbottom branching ratios can show also a strong
$\varphi_{A_b}$ dependence.
This strong phase dependence of CP-even observables like branching
ratios has to be taken into account in SUSY particle searches at
future colliders and the determination of the underlying MSSM parameters.
In order to estimate the expected accuracy in the determination of the
MSSM parameters we have made a global fit of masses, branching ratios
and production cross sections in two scenarios with small and large
$\tan\beta$. We have found that $A_t$ can be determined with an error
of 2 -- 3\,\%, whereas the error of $A_b$ is likely to be of the order
of 50 -- 100\,\%. 
Furthermore $\tan\beta$ can be determined with an error of 3\,\% and
the other fundamental MSSM parameters with errors of 1 -- 2\,\%.

However, in order to unambiguously establish CP violation in supersymmetry,
including the signs of the phases, the use of CP-odd
observables is inevitable.
We have studied T-odd asymmetries in neutralino and chargino
production with subsequent three-body decays, which are based on
triple product correlations between incoming and outgoing particles
and appear already at tree-level because of
spin correlations between production and decay.
The T-odd asymmetries can be as large as 20\,\% and will therefore be
an important tool for the search for CP violation in supersymmetry and
the unambiguous determination of the phases of the parameters
in the neutralino and chargino sectors.

\section*{Acknowledgements}

I thank A.~Bartl for many useful discussions.
This work has been supported by the European Community's Human Potential
Programme under contract HPRN-CT-2000-00149 ``Physics at Colliders''
and by the ``Fonds zur F\"orderung der wissenschaftlichen For\-schung''
of Austria, FWF Project No.~P16592-N02.

\end{document}